\documentclass[journal]{IEEEtran}
\usepackage{amsmath,amsfonts}
\usepackage{hyperref}
\usepackage{algorithm}
\usepackage{algpseudocode} 
\usepackage{array}
\usepackage[caption=false,font=normalsize,labelfont=sf,textfont=sf]{subfig}
\usepackage{textcomp}
\usepackage{stfloats}
\usepackage{url}
\usepackage{verbatim}
\usepackage{graphicx}
\usepackage{amsmath,amssymb}
\usepackage{xcolor}

\usepackage{cite}
\usepackage{amsthm}
\usepackage{marginnote}
\usepackage[normalem]{ulem}
\usepackage{multirow}

\newtheorem{thm}{Theorem}[section]

\newtheorem{lem}[thm]{Lemma}

\newcommand{\mrnote}[2]{#2}

\begin{document}	
	\title{Local monotone operator learning using non-monotone operators: MnM-MOL}
	\author{Maneesh John,~\IEEEmembership{Student Member,~IEEE,} Jyothi Rikhab Chand,  Mathews Jacob,~\IEEEmembership{Fellow,~IEEE}
		% <-this % stops a space
		\thanks{Maneesh John, Jyothi Rikhab Chand and Mathews Jacob are with the Department of Electrical and Computer Engineering at the University of Iowa, Iowa City, IA, 52242, USA (e-mail: maneesh-john@uiowa.edu; jyothi-rikhabchand@uiowa.edu; mathews-jacob@uiowa.edu). This work is supported by NIH grants R01-AG067078, R01-EB031169, and R01-EB019961.}}
	% stops a space
	
	% The paper headers
	\markboth{Journal of \LaTeX\ Class Files,~Vol.~14, No.~8, August~2021}%
	{Shell \MakeLowercase{\textit{et al.}}: A Sample Article Using IEEEtran.cls for IEEE Journals}
	
	\maketitle
	
	\begin{abstract}
The recovery of magnetic resonance (MR) images from  undersampled measurements is a key problem that has seen extensive research in recent years. Unrolled approaches, which rely on end-to-end training of convolutional neural network (CNN) blocks within iterative reconstruction algorithms, offer state-of-the-art performance. These algorithms require a large amount of memory during training, making them difficult to employ in high-dimensional applications. Deep equilibrium (DEQ) models and the recent monotone operator learning (MOL) approach were introduced to eliminate the need for unrolling, thus reducing the memory demand during training. Both approaches require a Lipschitz constraint on the network to ensure that the forward and backpropagation iterations converge. Unfortunately, the constraint often results in reduced performance compared to unrolled methods. The main focus of this work is to relax the constraint on the CNN block in two different ways. Inspired by convex-non-convex regularization strategies, we now impose the monotone constraint on the sum of the gradient of the data term and the CNN block, rather than constrain the CNN itself to be a monotone operator. This approach enables the CNN to learn possibly non-monotone score functions, which can translate to improved performance.  In addition, we only restrict the operator to be monotone in a local neighborhood around the image manifold. Our theoretical results show that the proposed algorithm is guaranteed to converge to the fixed point and that the solution is robust to input perturbations, provided that it is initialized close to the true solution. Our empirical results show that the relaxed constraints translate to improved performance and that the approach enjoys robustness to input perturbations similar to MOL. 
\end{abstract}

	\begin{IEEEkeywords}
		Model-based deep learning, Monotone operator learning, Deep equilibrium models.
	\end{IEEEkeywords}
	
	\section{Introduction}

\noindent The recovery of magnetic resonance (MR) images from undersampled measurements has been a subject of extensive research \cite{fessler2010model}. Deep learning (DL) algorithms \cite{gregor2010learning,romano2017little,hammernik2018learning,aggarwal2018modl,sun2019online,ryu2019plug,sun2021scalable,xiang2021fista,mol} have been introduced, which offer improved performance and faster inference compared to traditional compressed sensing (CS) methods \cite{lustig2007sparse}, making them attractive for a variety of clinical applications. Many of these DL schemes are iterative model-based approaches evolved from CS algorithms, where the proximal operator is replaced by a deep convolutional neural network (CNN) denoiser. The CNN may be pre-learned, as in plug-and-play methods \cite{romano2017little,sun2019online,ryu2019plug}. However, unrolling the iterative algorithm for a fixed number of steps, followed by end-to-end learning of the CNN blocks, is observed to offer improved performance \cite{hammernik2018learning,aggarwal2018modl}. A challenge with unrolled approaches is their high memory demand. Deep Equilibrium (DEQ) models use fixed-point iterations to eliminate the need for unrolling, thereby reducing the memory demand. 

Despite not being as powerful as their DL counterparts, CS methods have theoretical guarantees that include the uniqueness of the solution, robustness to input perturbations, and guaranteed convergence to the minimum of the cost function. The sensitivity of DL methods to input perturbations and model mismatches is debated; deep neural networks are reported to be more fragile to input perturbations than conventional algorithms \cite{antun2020instabilities}, while some recent work presents a more optimistic view \cite{darestani2021measuring, genzel2022solving}. Approaches to improve stability include heuristic approaches \cite{acid,mehmet}, and constraining the CNN to model convex functions \cite{unserconvex,amos2017icnn}. While the latter approach inherits many of the desirable properties of CS methods, the convexity constraint translates to poor performance. Another strategy is the recent monotone operator learning (MOL) approach, which generalizes the notion of iterative convex algorithms \cite{pesquet2021learning,mol} to DL. We note that the gradient of a convex penalty is a monotone operator. However, since arbitrary monotone functions need not be subdifferentials of convex functions, this approach is not equivalent to the convex DL approaches \cite{unserconvex,amos2017icnn} discussed above. The monotone constraint is enforced by representing the operator as a residual CNN, where the Lipschitz constant of the CNN block is restricted by using spectral normalization. These methods offer guaranteed uniqueness, convergence, and robustness to input perturbations. Empirical results show the good convergence and improved robustness of these models to input perturbation. However, the strict monotone constraint still translates to lower performance than unrolled algorithms. 

The focus of this paper is to introduce two novel extensions to further improve the performance of the MOL framework. Both of these extensions are aimed at relaxing the constraints on the CNN block. The first extension is motivated by the elegant convex-non-convex (CNC) variational framework \cite{parekh,lanza}.  CNC algorithms enable the use of non-convex priors to improve performance, while maintaining the favorable attributes of convex algorithms. Specifically, the non-convex penalty is carefully designed to ensure that the overall cost function, which is the sum of the quadratic data-consistency term and the non-convex prior, retains convexity. Motivated by the CNC approach, we propose to co-design the CNN with the forward model. In particular, we introduce a gradient descent algorithm to recover the image from the undersampled MRI measurements. Rather than constraining the gradient of the prior to be monotone, we only constrain the sum of the gradients of the data term and the prior to be monotone. This approach allows the CNN to learn potentially non-monotone functions, which may translate to improved performance. The proposed monotone-non-monotone MOL (MnM-MOL) scheme also has theoretical guarantees of uniqueness, convergence, and robustness similar to MOL, while the relaxation of the constraint translates to improved performance. 

We further extend the MOL scheme by relaxing the global monotone constraint assumed in \cite{mol}. Rather than constraining the operator to be monotone across its entire domain, we introduce a weaker local monotone property that must be satisfied within a ball of radius $\delta$ around each training data point in the data manifold $\mathcal M$. We note that the theoretical guarantees in \cite{mol}, which assume global monotone operators, are not valid in our setting with weaker local monotone constraints. We show that the fixed point of the gradient descent algorithm is unique within the ball when the local monotone condition is satisfied. We introduce novel convergence guarantees for the algorithm, provided that the algorithm is initialized within a ball of radius $\delta$ centered at the true solution. We propose to initialize the algorithm using fast least squares approaches (e.g. SENSE\cite{pruessmann1999sense} in MR imaging), which is sufficient to guarantee convergence to the unique fixed point within the ball. Our theoretical results also show that the proposed approach is robust to small input perturbations.  Comparison of our proposed MnM-MOL algorithm with the MOL scheme shows an improvement in performance, while preserving the practical benefits of MOL. 

\section{Background}
\subsection{MR image reconstruction}
We consider recovery of an image $\mathbf x \in \mathbb{C}^{m}$ from its noisy undersampled measurements $\mathbf b \in \mathbb{C}^{n}$, specified by
\begin{equation}
\label{eq:linear}
\mathbf b = \mathbf A\mathbf x+ \mathbf n,
\end{equation}
where $\mathbf A \in \mathbb{C}^{n \times m}$ is a linear operator and $\mathbf n \sim \mathcal N(\mathbf 0, \sigma^2 \mathbf I)$ is additive white Gaussian noise. The conditional probability of the measurements is specified by the Gaussian distribution: $p(\mathbf b|\mathbf x) = \mathcal N(\mathbf A\mathbf x, \sigma^2 \mathbf I)$. The maximum-a-posteriori (MAP) estimation of $\mathbf x$ from the measurements $\mathbf b$ poses the recovery as the maximum of the posterior $p(\mathbf x|\mathbf b)$. Using Bayes' rule, $p(\mathbf x|\mathbf b) \propto p(\mathbf b|\mathbf x)p(\mathbf x)$ we have 
\begin{equation}
\label{eqn:cs}
\mathbf x_{\rm MAP} = \arg \min_x \underbrace{\frac{\lambda}{2}\|\mathbf A\mathbf x - \mathbf b\|_2^2}_{\mathcal D(\mathbf x) = -\log~p(\mathbf b|\mathbf x)} + \underbrace{\phi(\mathbf x)}_{-\log p(\mathbf x)}. 
\end{equation}
Here, $\lambda = \frac{1}{\sigma^2}$. The first term $\mathcal D(\mathbf x) = -\log~p(\mathbf b|\mathbf x)$ is the data consistency term, while the second term is the log-prior. CS algorithms use strongly convex priors (e.g., $\phi(\mathbf x) = \|\mathbf x\|_{1}$) to result in a convex cost function with a unique minimum. When $\phi$ is strongly convex, $\mathbf x_{\rm MAP}$ is unique. 

We note that the minimum of \eqref{eqn:cs} satisfies the fixed-point relation:
\begin{equation}
\label{eqn:grad}
\lambda~\mathbf A^H (\mathbf A\mathbf x - \mathbf b) + \underbrace{\nabla_{\mathbf x}~\phi(\mathbf x)}_{\mathcal F(\mathbf x)} = 0.
\end{equation}
Here, $\mathbf A^H$ is the Hermitian operator of $\mathbf A$ and $\nabla_{\mathbf x}\phi(\mathbf x) = -\nabla_{\mathbf x}\log p(\mathbf x)$ is often termed the score function. In recent years, several researchers have proposed to model the score using a CNN $\mathcal F_{\theta}$ with weights denoted by $\theta$ \cite{mol}.

\subsection{Monotone operator learning \cite{mol}}
 When $\phi(\mathbf x)$ is a closed proper $m$-convex function, the operator $\mathcal F_{\theta}$ is $m$-monotone: 
\begin{equation}
\label{eqn:monotonicity_definition}
\Re \left(\Big\langle \mathbf x_1 - \mathbf x_2, \mathcal F_{\theta}(\mathbf x_1) - \mathcal F_{\theta}(\mathbf x_2) \Big\rangle\right) \geq m\|\mathbf x_1- \mathbf x_2\|_2^2, \hspace{6pt} m>0,
\end{equation}
for all $\mathbf x_1, \mathbf x_2 \in \mathbb{C}^m$. An example that is easy to appreciate is the case of linear operators, when  \eqref{eqn:monotonicity_definition} implies that $\left\langle\mathbf x, \mathbf F\mathbf x\right\rangle\geq m\|\mathbf x\|^2$, or equivalently $\mathbf F$ is strictly positive definite. 
 We note that an $m$-monotone function $\mathcal F_{\theta}$ need not be the subdifferential of a convex function; MOL  algorithms \cite{pesquet2021learning,mol} that constrain $\mathcal F_{\theta}$ to be m-monotone are not equivalent to convex deep learning methods  \cite{unserconvex,amos2017icnn}. The following result was used to constrain $\mathcal F_{\theta}$ as a monotone operator in \cite{mol}: \begin{lem}\cite{mol}
 \label{residual}
$\mathcal F_{\theta}= \mathcal I -\mathcal H_{\theta}$ is $m$-monotone with $0<m<1$ if the Lipschitz constant of $\mathcal H_{\theta}$ is $L=1-m$:
\begin{equation}
\label{globalmonotone}
    \|\mathcal H_{\theta}(\mathbf x_1)-\mathcal H_{\theta}(\mathbf x_2)\|_{2} \leq (1-m)~\|\mathbf x_1-\mathbf x_2\|_{2}; ~~\forall \mathbf x_1, \mathbf x_2 \in \mathbb{C}^m.
\end{equation}
\end{lem}
\noindent In particular, $\mathcal F_{\theta}=\mathcal I-\mathcal H_{\theta}$ was designed as a residual network that involved the CNN $\mathcal H_{\theta}$. Spectral normalization of the CNN layers can guarantee the global Lipschitz constant of $\mathcal H_{\theta}$ is less than $1-m$ and hence ensure that $\mathcal F_{\theta}$ is $m-$monotone. When $\mathcal F_{\theta}$ is monotone, the fixed point of \eqref{eqn:grad} is unique \cite{mol}. A forward-backward splitting algorithm:
\begin{eqnarray}
\label{pnp-mol}
\mathbf x_{k+1} &=& \underbrace{(\mathcal I + \alpha \mathbf A^H\mathbf A)^{-1}(\mathcal I - \alpha \mathcal F_{\theta})}_{\mathcal T}(\mathbf x_k),
\end{eqnarray}
which is guaranteed to converge to the unique fixed point was introduced in \cite{mol}. The monotone property also guarantees the robustness of the algorithm to input perturbations \cite{mol}. 
 
The operator $\mathcal F_{\theta}$ can be prelearned as in plug-and-play methods \cite{venkatakrishnan2013plug,buzzard2018plug,romano2017little,sun2019online,ryu2019plug,sun2021scalable} or learned end-to-end \cite{gregor2010learning,hammernik2018learning,aggarwal2018modl, xiang2021fista,mol}. The deep equilibrium framework \cite{gilton2021deep,bai2019deep} was used in \cite{mol} for memory efficient end-to-end learning. DEQ models assume that the forward iterations in \eqref{pnp-mol} are run until convergence to the fixed point $\mathbf x^*$ which satisfies $\mathbf x^*=\mathcal T(\mathbf x^*,\theta)$. This approach allows for computation of the backpropagation steps using fixed point iterations \cite{bai2019deep,gilton2021deep} with just one physical CNN block. 

The empirical results in \cite{mol} show that the MOL algorithm offers faster convergence and better robustness to input perturbations than other DL solutions. However, these gains come at the cost of performance, which is lower than that of unrolled methods. The main cause is the global Lipschitz constraint on $\mathcal H_{\theta}$, which unrolled methods do not require. This reduction in performance is not surprising, considering that non-convex priors are reported to yield improved results over convex approaches in the context of CS. Although weaker local monotone conditions improved performance, the theoretical guarantees in \cite{mol} are invalid in this setting. 

\section{Proposed approach}

In this paper, we present two novel relaxations aimed at enhancing the effectiveness of the MOL scheme. The first approach is inspired by CNC methods, which customize the non-convex regularizer $\phi(\mathbf x)$ to align with the data-term, resulting in a convex combined cost function. We propose to co-learn a possibly non-monotone operator $\mathcal F_{\theta}$ with the gradient of the data consistency term in Section \ref{global}. We then introduce a relaxation of the global monotone constraint into a local monotone constraint in Section \ref{local}. Both of these relaxations are expected to improve performance. 

\subsection{MOL with Non-Monotone Operators (MnM-MOL) }
\label{global}
We propose to learn a custom non-monotone operator $\mathcal F_{\theta}$ such that:
\begin{eqnarray}\label{relax}
\mathcal Q_{\theta}(\mathbf x) = 	\mathbf A^H\mathbf A\mathbf x + \frac{1}{\lambda}\mathcal F_{\theta}(\mathbf x)
\end{eqnarray}
is monotone. This is a relaxation of the approach in \cite{mol}, where $\mathcal F_{\theta}$ itself is required to be a monotone operator. We note that if $\mathcal F_{\theta}$ is $m$-monotone, $\mathcal Q_{\theta}$ is guaranteed to be $m$-monotone, irrespective of the choice of $\mathbf A$. However, the relaxation allows $\mathcal F_{\theta}$ to learn possibly non-monotone score functions, which can result in improved performance \footnote{For example, when $\mathcal F_{\theta}=\mathbf F$ is linear and $\mathbf A$ is of full rank, it is possible to have a negative definite $\mathbf F$ (e.g. $\mathbf{F} = \beta\mathbf{I} - \lambda~\mathbf{A}^H \mathbf{A}$ and $\beta << \lambda$), such that $\mathcal Q_{\theta}$ is still positive definite or monotone.}.

We introduce a steepest descent algorithm to converge to the fixed point \eqref{eqn:grad} denoted by $\mathbf x^*$:
\begin{eqnarray}\nonumber
     \mathbf x_{k+1} &=& \mathbf x_k - \gamma(\lambda (\mathbf A^H\mathbf A\mathbf x_k-\mathbf A^H\mathbf b) + \mathcal F_{\theta}(\mathbf x_k)) \\\label{iterative}
     &=& \mathbf x_k - \gamma\lambda ~\mathcal Q_{\theta}(\mathbf x_k) + \gamma\lambda\; \mathbf A^H\mathbf b%\\\nonumber
     %&=& \mathbf x_k \left(1-\gamma\lambda\right) + \gamma\lambda\left(\mathbf x_k - ~\mathcal Q_{\theta}(\mathbf x_k)\right) + \gamma\lambda\; \mathbf A^T \mathbf b
\end{eqnarray}
Here, $\gamma$ is the step-size. When $\gamma=1/\lambda$, the algorithm simplifies to: \begin{equation}\label{gd_simple}
\mathbf x_{k+1} = \underbrace{\mathbf x_k-\mathbf A^H\mathbf A\mathbf x_k - \frac{1}{\lambda}\mathcal F_{\theta}(\mathbf x_k)}_{\mathcal H_{\theta}(\mathbf x_k)=(\mathcal I-\mathcal Q_{\theta})(\mathbf x_k)} + \mathbf A^H \mathbf b.
\end{equation} 
We note that the fixed point of this algorithm satisfies \eqref{eqn:grad}.  A global monotone constraint on $\mathcal Q_{\theta}$ is sufficient to guarantee that the fixed point of the above algorithm satisfies all the desirable properties of the MOL scheme in \cite{mol}. The following result shows that the monotone property of $\mathcal Q_{\theta}$ guarantees the uniqueness of the fixed point of \eqref{eqn:grad}.
\begin{lem}
\label{lem0}
    The fixed point of \eqref{gd_simple} is unique when $\mathcal Q_{\theta}$ is $m> 0$ monotone.
\end{lem}
The proof is included in Appendix A and is along the same lines as in \cite{mol}. Similarly, the convergence and robustness guarantees in \cite{mol} can be translated into this setting. 

We note from Lemma \ref{residual} that $\mathcal Q_{\theta}$ is $m$-monotone if the Lipschitz constant of its residual $\mathcal H_{\theta} = \mathcal I - \mathcal Q_{\theta}$ is restricted. Because $\mathcal Q_{\theta}$ is not a CNN as in \cite{mol}, the global Lipschitz constraint cannot be enforced by using spectral normalization \cite{miyato2018spectral}. In particular, $\mathcal Q_{\theta}$, which is specified by \eqref{relax}, is dependent on the CNN $\mathcal F_{\theta}$ as well as the forward operator $\mathbf A$. Due to this challenge, and in the interest of relaxing the global Lipschitz condition in \cite{mol}, we now introduce a local condition. 

\subsection{Locally monotone operators}
\label{local}

\begin{figure}
    \centering
    \includegraphics[width=0.45\textwidth]{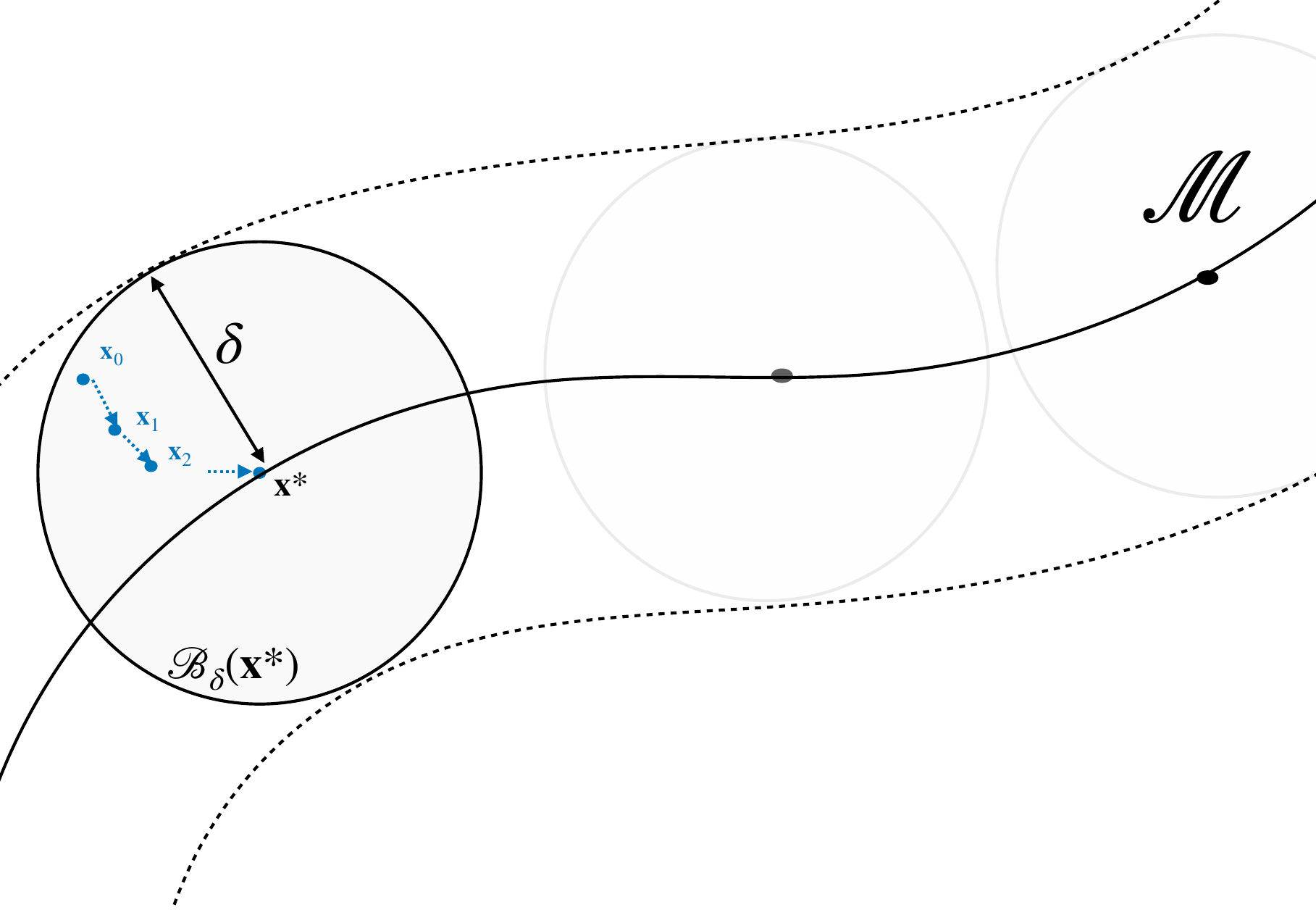}
    \caption{Illustration of the local monotone condition: we restrict the local Lipschitz constant of $\mathcal H_{\theta}$ in $\mathcal B_{\delta}(\mathbf x)$, which is a ball of radius $\delta$ centered at $\mathbf x$ around each of the training samples in the data manifold $\mathcal M$ as specified by \eqref{Lipschitz}. This training strategy ensures that $\mathcal Q_{\theta}$ is $m$-monotone in a tube of radius $\delta$ around the data manifold $\mathcal M$, when the manifold is well-sampled. Our results guarantee the convergence of the algorithm to fixed point  $\mathbf x^*\in \mathcal M$, provided the algorithm is initialized with $\mathbf x_0$, restricted to $\mathcal B_{\delta}(\mathbf x^*)$. Similarly, the algorithm is robust to input perturbations satisfying \eqref{noisebound} }
    \label{manifold}
\end{figure}

We now concentrate on constraining the operators to be locally monotone in proximity to the image manifold, as illustrated in Fig. \ref{manifold}. We define an operator $\mathcal Q_{\theta}$ to be locally $m>0$ monotone at $\mathbf z$ if it satisfies
\begin{eqnarray}\nonumber
\Re\Big\langle \mathbf z_2-\mathbf z_1, \mathcal \mathcal Q_{\theta}(\mathbf z_2)-\mathcal Q_{\theta}(\mathbf z_1) \Big\rangle&&\geq ~ m~\|\mathbf z_2 - \mathbf z_1\|^2,\\\label{weaker}&&\forall \mathbf z_1,\mathbf z_2 \in \mathcal B_{\delta}(\mathbf z),
\end{eqnarray}
where
\begin{equation}
\label{localmonotonedefinition}
 \mathcal B_{\delta}(\mathbf z) = \{\tilde{\mathbf {z}}:~\|\tilde{\mathbf {z}}-\mathbf z\|\leq \delta\},   
\end{equation}
is a ball of radius $\delta$ centered at $\mathbf z$. If the above relation holds for all $\mathbf x(k)$ in the image manifold, $\mathcal Q_{\theta}$ will be locally monotone in a tube around the manifold as shown in Fig. \ref{manifold}. As $\delta \rightarrow \infty$, the local monotone condition will be equivalent to the global monotone condition in \eqref{eqn:monotonicity_definition}. 

The following result provides a sufficient condition on $\mathcal H_{\theta} = \mathcal I-\mathcal Q_{\theta}$ in \eqref{gd_simple}, which will guarantee the local monotone property. 
\begin{lem}
\label{lem1}
The operator $\mathcal Q_{\theta} = \mathcal I-\mathcal H_{\theta}$ is locally monotone in $\mathcal B_{\delta}(\mathbf x)$ if the local Lipschitz constant of $\mathcal H_{\theta}$ within $\mathcal B_{\delta}(\mathbf x^*)$ is $1-m$:     
\begin{equation}
\label{localmonotone}
    \max_{\mathbf z_1,\mathbf z_2\in B_{\delta}(\mathbf x)} \|\mathcal H_{\theta}(\mathbf z_2)-\mathcal H_{\theta}(\mathbf z_1)\| \leq (1-m)~\|\mathbf z_2-\mathbf z_1\|.
\end{equation}
\end{lem}
The proof can be found in Appendix B. We note that \eqref{localmonotone} is a weaker version of \eqref{globalmonotone}, similar to \eqref{weaker} being a weaker local version of \eqref{eqn:monotonicity_definition}.

\subsection{Theoretical guarantees}
We will now introduce novel theoretical guarantees on the uniqueness, convergence, and robustness of the steepest descent algorithm in \eqref{gd_simple}. Unlike the results in \cite{mol} which assume the global monotone condition specified by \eqref{globalmonotone}, these results only require the weaker local monotone condition specified by \eqref{localmonotone}. The following result shows that the locally monotone property of $\mathcal Q_{\theta}$ guarantees the uniqueness of the fixed point $\mathbf x^*\in \mathcal M$ of \eqref{eqn:grad} within $\mathcal B_{\delta}(\mathbf x^*)$. 

\begin{lem}
\label{lem3}
    Assume that $\mathbf x^*$ is a fixed point of \eqref{eqn:grad}. If $\mathcal Q_{\theta}$ is locally $m$-monotone within $\mathcal B_{\delta}(\mathbf x^*)$, then \eqref{eqn:grad} does not have any other fixed point within $\mathcal B_{\delta}(\mathbf x^*)$.
\end{lem}

\begin{figure}
    \centering
    \includegraphics[width=0.47\textwidth,keepaspectratio=true,trim={0cm 11cm 5.5cm 0cm},clip]{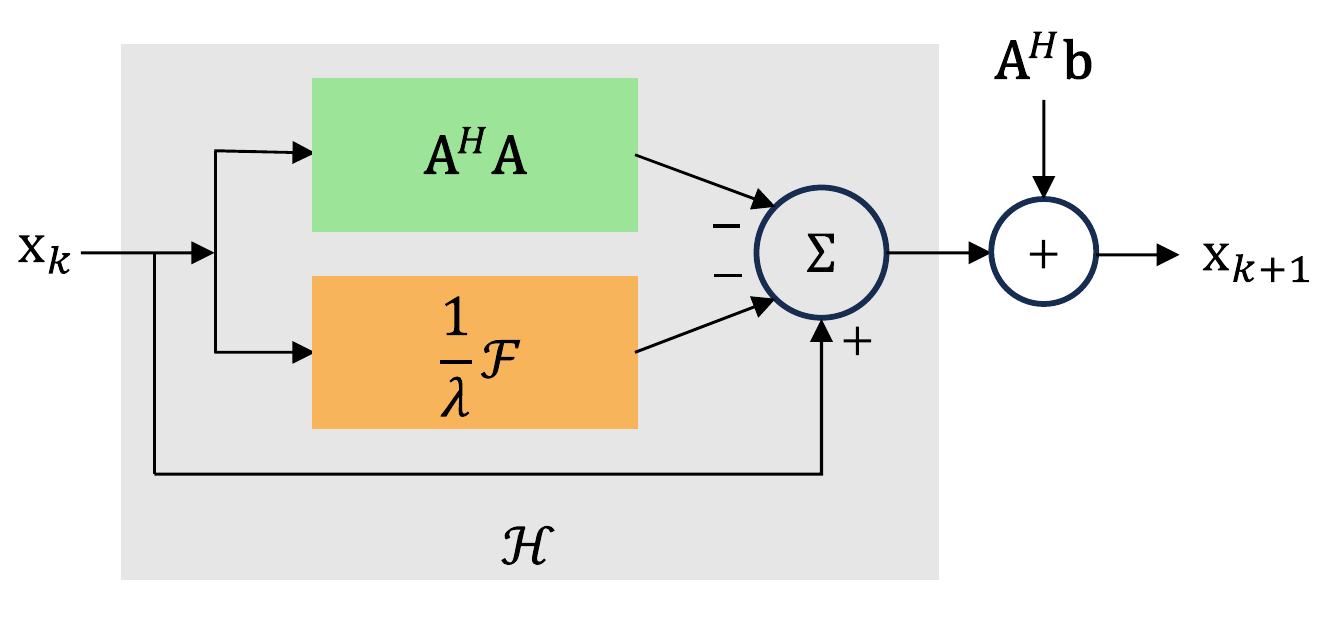}
    \caption{One step of the iterative algorithm, specified by \eqref{gd_simple}.  Lemma \ref{lem3} guarantees that  \ref{eqn:grad} has a unique solution within $\mathcal B_{\delta}(\mathbf x^*)$. The algorithm is initialized by $\mathbf x_0 \in \mathcal B_{\delta}(\mathbf x^*)$ and iterated until convergence. Lemma \ref{lem2} guarantees the convergence of the algorithm to the unique fixed point $\mathbf x^*$ in $\mathcal B_{\delta}(\mathbf x)$, provided it is initialized within $\mathcal B_{\delta}(\mathbf x)$.  Lemma \ref{lem4} guarantees that the solution is robust to  perturbations in $\mathbf b$.  }
    \label{fig:enter-label}
\end{figure}

Details of the proof can be found in Appendix C. The isolated nature of the fixed point enables us to derive convergence guarantees. We will now show that the steepest descent algorithm in \eqref{gd_simple} converges to the unique fixed point in $\mathcal B_{\delta}(\mathbf x^*)$, when initialized with a point $\mathbf x_0 \in \mathcal B_{\delta}(\mathbf x^*)$. We note from Lemma \ref{lem3} that there is no other fixed point within $\mathcal B_{\delta}(\mathbf x^*)$. 

\begin{lem}
\label{lem2}
Let the iterative algorithm specified by \eqref{iterative} be initialized with $\mathbf{x_0} \in \mathcal B_{\delta}(\mathbf x^*)$. Then the iterations converge to the fixed point $\mathbf x^* $ as: 
\begin{eqnarray*}
     \|\mathbf x_{k}-\mathbf x^*\|_2 &\leq& 
\|\mathbf x_0 -\mathbf x^*\|_2~\Big(1-\gamma \lambda m\Big)^k
\end{eqnarray*}
whenever $\gamma \leq 1/\lambda$ and when $\mathcal H_{\theta} = \mathcal I-\mathcal Q_{\theta}$ satisfies \eqref{localmonotone} within $\mathcal B_{\delta}(\mathbf x^*)$.
\end{lem}
Proof can be found in Appendix D. We note the special case $\gamma\lambda =1$ offers the fastest convergence. We hence restrict ourselves to this setting for simplicity in the rest of the paper. We note that the algorithm must be initialized within $\mathcal B_{\delta}(\mathbf x)$ for convergence. We discuss initialization strategies that satisfy this constraint in Section \ref{initialization}.

 We will now analyze the sensitivity of the solutions of the iterative algorithm, when the measurements $\mathbf b$ are corrupted by small perturbations.
\begin{lem}
\label{lem4}
Assume that $\mathbf x^* \in \mathcal M$ and $\mathbf y^*$ are fixed points that correspond to the measurements $\mathbf b$ and perturbed measurements $\mathbf b+\mathbf n$, respectively, where
\begin{equation}
\label{noisebound}
\|\mathbf A^H{\mathbf n}\|_2 \leq m ~\delta.
\end{equation}
When \eqref{localmonotone} is satisfied, the norm of the perturbation in the fixed point $\|\Delta\|_2 = \|\mathbf y^*-\mathbf x^*\|_2$ is bounded by 
    \begin{equation}
\|\Delta\|\leq \frac{\|\mathbf A^H\mathbf n\|_2}{m}
\end{equation}
\end{lem}
The above relation shows that the norm of the perturbation in the solution is bounded by the norm of the input perturbation, multiplied by $1/m$. A higher value of $m$ leads to a more robust algorithm. We also note from \eqref{noisebound} that the above relation is only valid when the input perturbations are small. The bound ensures that the iterations and the final solution remain within $\mathcal B_{\delta}(\mathbf x^*)$; perturbations larger than \eqref{noisebound} may push the algorithm out of the convergence basin. 

\subsection{Imposing local monotonicity during training}
We are motivated by \cite{bungert2021clip} to train the MOL algorithm using a training loss that minimizes a constrained optimization problem. This approach is similar to \cite{pesquet2021learning}, which employed Jacobian regularization. We estimate the local Lipschitz constant within $\mathcal B_{\delta}(\mathbf x^*)$ by solving the maximization problem:
\begin{equation}
	\label{Lipschitz}
	L\left[\mathcal H_{\theta}(\mathbf x^*)\right] = \max_{\mathbf z_2,\mathbf z_1 \in \mathcal B_{\delta}(\mathbf x^*)}\frac{\|\mathcal H_{\theta}(\mathbf z_2) -\mathcal H_{\theta}(\mathbf z_1)\|_2}{\|\mathbf z_2-\mathbf z_1\|_2}
\end{equation} 
We identify the most unfavorable $\mathbf z_1$ and $\mathbf z_2$ by utilizing projected gradient ascent, beginning with a random initialization within $\mathcal B_{\delta}(\mathbf x^*)$. Specifically, we update $\mathbf z_1$ and $\mathbf z_2$ to maximize the cost function of \eqref{Lipschitz} by means of gradient ascent, and subsequently project them to $\mathcal B_{\delta}(\mathbf x^*)$ at each step.

\subsection{End to end training of MnM-MOL network}

In the supervised learning setting, we propose to solve: 
\begin{eqnarray}\nonumber
	\label{supervisedloss}
	\theta^* &=& \arg \min_{\theta} \sum_{k=0}^{N_t} \|\mathbf x^*(k,\theta) - \mathbf x(k)\|_2^2\\\nonumber&&\qquad~\mbox{such that }~L\left[\mathcal H_{\theta}\left(\mathbf x^*(k)\right)\right]\leq T; k=0,..,N_t\\
\end{eqnarray}
The above loss function is minimized w.r.t. weights $\theta$ of the CNN $\mathcal H_{\theta}$. The ground truth images in the training dataset and the corresponding under-sampled measurements are denoted as $\mathbf x(k)$ and $\mathbf b(k); k=0,..,N_t$, respectively. The fixed point for the $k^{\rm th}$ sample is denoted as $\mathbf x^*(k)$.
Note that the fixed point of \eqref{iterative} $\mathbf x^*(k,\theta)$ is dependent on the CNN parameters $\theta$. Within each iteration, we determine $L\left[\mathcal H_{\theta}(\mathbf x^*(k))\right]$ in \eqref{Lipschitz} using projected gradient ascent \footnote{The local Lipschitz constant in (\ref{supervisedloss}) is computed for a fixed $\theta$.}. The threshold for the Lipschitz constant is selected as $T = 1-m$. We solve the above constrained optimization scheme by adding a loss term:

\begin{equation}
	\label{finalloss}
	\min_{\theta}~\sum_{k=0}^{N_t} \underbrace{\|\mathbf x^*(k) - \mathbf x(k)\|_2^2 +  \beta ~{\rm ReLU}\big(L\left[\mathcal H_{\theta}(\mathbf x^*(k))\right]-T\big)^2}_{C_k} .
\end{equation}

Note that ${\rm ReLU}\big(L\left[\mathcal H_{\theta}(\mathbf x^*(k))\right]-T\big)$ is zero if the constraint is met. In this work, we set $\beta=1$, which was sufficient to enforce the constraint $L\left[\mathcal H_{\theta}(\mathbf x)\right]<T; \forall \mathbf x\in \mathcal M$.

In each iteration, we determine $L\left[\mathcal H_{\theta}(\mathbf x^*(k)\right]$ for each training data point. In particular, we determine the worst-case perturbations $\mathbf z_1^*,\mathbf z_2^* \in \mathcal B_{\delta}(\mathbf x^*(k))$ for each fixed point $\mathbf x^*(k)$ by maximizing the cost function of \eqref{Lipschitz} using projected gradient ascent. The optimal $\mathbf z_1^*,\mathbf z_2^* \in \mathcal B_{\delta}(\mathbf x^*(k))$ is then used to compute $L\left[\mathcal H_{\theta}(\mathbf x^*(k))\right]$ in \eqref{supervisedloss} using \eqref{Lipschitz}.

The training algorithm is summarized by the pseudocode shown in Algorithm \ref{training}, which is illustrated for a batch size of a single image and gradient descent for simplicity.
\begin{algorithm}
	\caption{: Input: training data $\mathbf x(k),\mathbf b(k);k=1,..,N_t$}
	 \label{training}
	\mrnote{R1.4}{\begin{algorithmic}[1]
		\For {ep = $1,2,\ldots$ }
			\For {$k=1,2,\ldots, N_t$ }
				\State Determine $\mathbf x^*(k,\theta)$ using DEQ forward iterations
				\State Estimate $L[\mathcal{H_\theta}(\mathbf x^*(k))]$ using \eqref{Lipschitz}
				\State $C_i = \|\mathbf x^*(k)-\mathbf x(k)\|_2^2 + \beta ~{\rm ReLU}(L[\mathcal{H_\theta}]-T)^2\ $
            \State Determine $\nabla_{\theta} C_i$ using DEQ backward iterations
    \State $\theta \leftarrow \theta - \gamma \nabla_{\theta} C_i$, where $C_i$ is the loss in \eqref{finalloss}
			\EndFor
		\EndFor
	\end{algorithmic} }
\end{algorithm}

\begin{figure}[!b]
	\centering
	\includegraphics[width=0.5\textwidth,keepaspectratio=true,trim={0cm 6cm 1cm 5.5cm},clip]{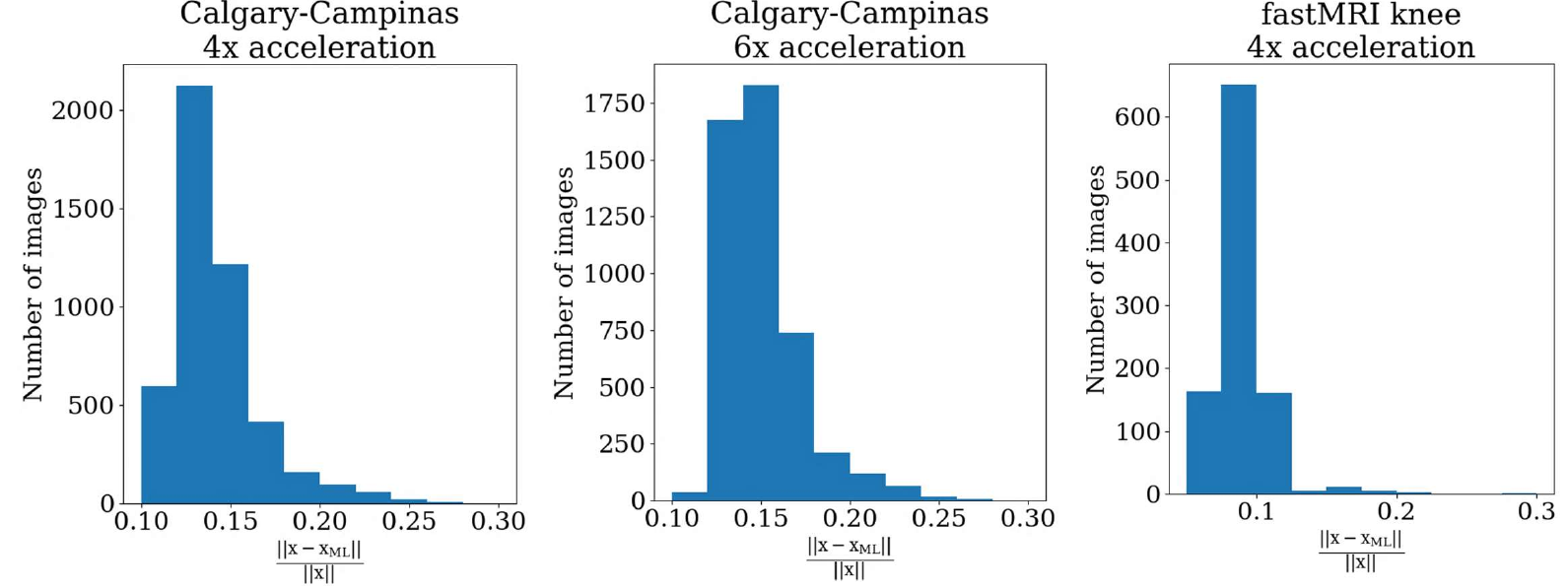}
	\caption{We initialize the proposed algorithm with the least-squares (SENSE) reconstruction scheme, which is fast to compute. The histogram of the norm of the difference between the reference and the least-squares solution, divided by the norm of the ground truth image, is shown above for different datasets and acceleration factors. The worst-case value was computed for each dataset at each acceleration to determine the $\delta$ value to train the proposed scheme. This strategy guarantees that the initialization $\mathbf x_0 \in \mathcal{B}_{\delta}(\mathbf x)$, which is needed for convergence. }
	\label{fig:histogram}
\end{figure}

\section{Experiments \& Results}
\subsection{2D Brain and knee datasets}

We used 2D multicoil brain data from the publicly available Calgary-Campinas Public (CCP) dataset \cite{souza2018open}. This data set consists of T1-weighted multi-coil brain scans from 117 healthy subjects, which were collected on a 3.0 Tesla MRI scanner. The scan parameters were TR (repetition time)/TE (echo time)/TI (inversion time) = 6.3 ms/2.6ms/650 ms or TR/TE/TI = 7.4ms/3.1ms/400ms. 
For the experiments, we selected a subset with fully sampled data (67 of 117) and divided them into training (47) and testing (20) sets. The k-space measurements were retrospectively undersampled along the phase and slice encoding directions using a four-fold or six-fold 2D nonuniform variable density mask.

We also validated our approach using multichannel knee data from the fastMRI challenge \cite{zbontar2018fastmri}. This data set includes 15-coil coronal proton-density weighted knee images with or without fat suppression. The sequence parameters are: matrix size 320 $\times$ 320, in-plane resolution 0.5mm $\times$ 0.5mm, slice thickness 3mm, TR ranging from 2200 to 3000 ms, and TE between 27 and 34 ms. We used the k-space measurements from 50 subjects for training and 10 for testing, respectively. The k-space measurements were retrospectively undersampled along the phase encoding direction using a four-fold 1D nonuniform variable density mask.

\subsection{Choice of $\delta$ and $m$}\label{initialization}
We note from Lemma \ref{lem2} that the algorithm must be initialized within $\mathcal B_{\delta}(\mathbf x^*)$ for it to converge to the fixed point $\mathbf x^*$. A higher value of $\delta$ will translate to a larger basin of attraction. However, we see from \eqref{localmonotone} that the local Lipschitz bound for $\mathcal H_{\theta}$ needs to be satisfied in a larger region, which  will result in lower performance. For instance, $\delta \rightarrow \infty$ would offer global convergence, but the global Lipschitz constraint translates into lower performance, as seen from \cite{mol}. Therefore, we choose the smallest $\delta$ that will satisfy convergence. 

We propose to initialize the algorithm using the least squares solution (known as SENSE \cite{pruessmann1999sense} in the MRI context), which is fast to compute. We search over all the training samples to determine the worst-case deviation of the least-square solutions from the reference images. We then choose $\delta$ such that the SENSE solutions are within $\mathcal B_{\delta}(\mathbf x^*)$ for all training data samples. We show the histogram of $\|\mathbf x_{LS}-\mathbf x^*\|_2/\|\mathbf x^*\|_2$ for the datasets in {Fig. \ref{fig:histogram}}. Based on these results, we propose to choose $\delta$ as the maximum value of $\|\mathbf x_{LS}-\mathbf x^*\|_2/\|\mathbf x^*\|_2$. 

\begin{table}[ht!]
\fontsize{7}{10}
\selectfont
\centering
\begin{tabular}{|c|cc|cc|}
\hline
& \multicolumn{2}{c|}{Four-fold Brain MRI} & \multicolumn{2}{c|}{Six-fold Brain MRI} \\ \hline
Methods & PSNR & SSIM & PSNR & SSIM \\ \hline 
SENSE  & 30.59 ± 1.47 & 0.904 ± 0.024 & 30.01 ± 1.35 & 0.894 ± 0.024 \\
MoDL   & 34.98 ± 1.79 & 0.968 ± 0.016 & 34.09 ± 1.72 & 0.963 ± 0.017 \\ 
MOL-SN  & 32.31 ± 1.54 & 0.940 ± 0.020 & 31.57 ± 1.48 & 0.929 ± 0.022 \\
MOL-L & 33.84 ± 1.55 & 0.961 ± 0.017 & 33.04 ± 1.63 & 0.952 ± 0.020 \\
\textbf{MnM-MOL} & \textbf{34.71 ± 1.71} & \textbf{0.965 ± 0.014} & \textbf{33.80 ± 1.74} & \textbf{0.959 ± 0.019} \\ 
\hline
\end{tabular}
\vspace{1em}
\caption{Quantitative comparisons on Calgary-Campinas brain dataset. Mean PSNR (dB) and mean SSIM are reported.}
\label{tab:calgary_perf_comp_tab} 
\end{table}

\begin{figure*}[!t]
	\centering
	\includegraphics[width=1.0\textwidth,keepaspectratio=true,trim={2cm 2.75cm 2cm 2.75cm},clip]{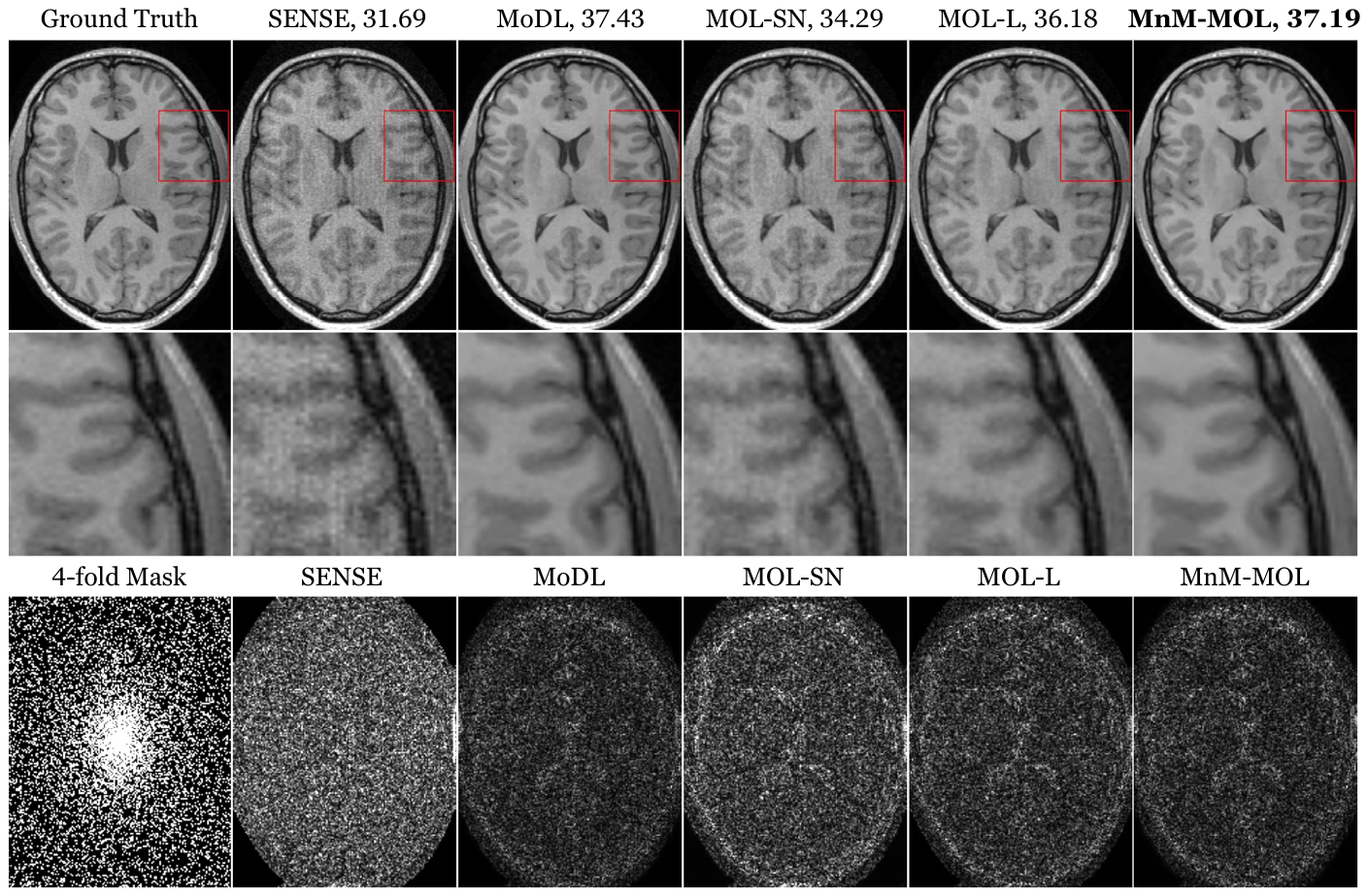}
	\caption{Reconstruction results of 4x accelerated multichannel brain data. PSNR values (dB) for each case are reported. The image in the first row of the first column was undersampled using a Cartesian 2D nonuniform variable density mask, as seen in the second row of the first column. The upper row displays the reconstructions (magnitude images), while the lower row shows the corresponding error images, where the intensity is scaled by 10x for better visualization. The quality of the MnM-MOL reconstructions is observed to be similar to that of the unrolled MoDL method in terms of PSNR. Visually, the MnM-MOL results appear to be less noisy and cleaner than those of MoDL. We also note that the performance of the MOL-SN scheme is worse than MOL-L, which we attribute to the strong constraints on the CNN.} 
	\label{fig:calgary-performance}
\end{figure*}

\subsection{Performance comparison in the parallel MRI setting}

The performance of multiple algorithms is compared on the two datasets described above. We compare the proposed MnM-MOL scheme with four other algorithms: the classical SENSE which is used as the initialization for MnM-MOL, an unrolled algorithm MoDL \cite{aggarwal2018modl} that is also trained end-to-end, the MOL scheme with a global monotone constraint enforced using spectral normalization, and the MOL scheme using the local monotone constraint. The regularization parameters of the $\ell_2$ SENSE approach are tuned to obtain the best performance, while the parameters of the remaining end-to-end methods are learned from the training data.

\begin{figure*}[!t]
	\centering
	\includegraphics[width=1.0\textwidth,keepaspectratio=true,trim={1.7cm 4cm 1.7cm 4cm},clip]{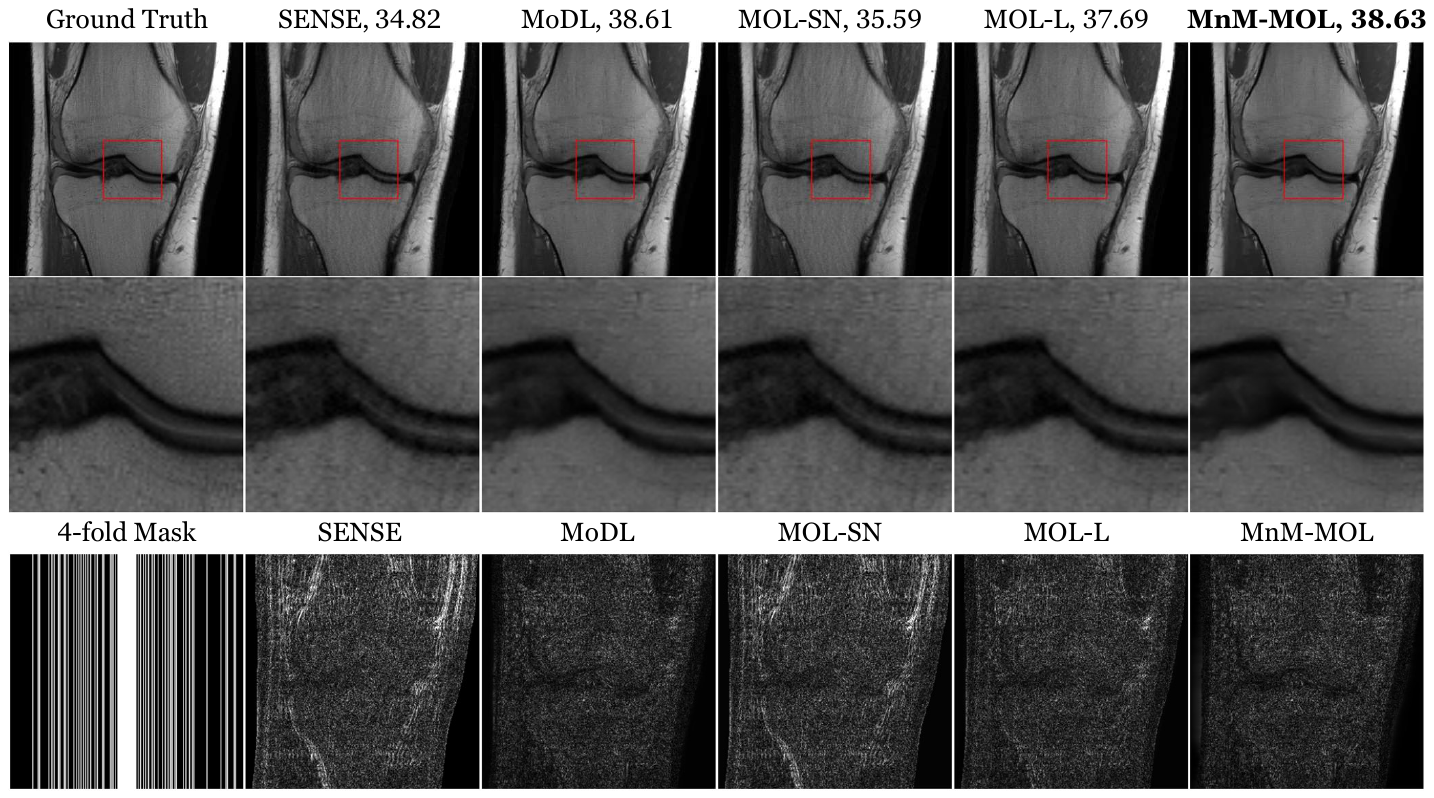}
	\caption{Reconstruction results of 4x accelerated multichannel fastMRI knee data with variable density sampling. PSNR values (dB) are reported for each case. The image in the first row of the first column was undersampled along the phase-encoding direction using a Cartesian 1D non-uniform variable density mask, as shown in the second row of the first column. The top row shows reconstructions (magnitude images), while the bottom row shows the corresponding error images. We note that the quality of the MnM-MOL reconstructions is comparable to that of the unrolled method MoDL. Both MOL-SN and MOL-L show lower performance. The improved performance of MnM-MOL over other MOL schemes can be attributed to the relaxed constraints on the CNN.}
	\label{fig:fastmri-performance}
\end{figure*}

\begin{table}[t!]
\fontsize{7}{10}
\selectfont
\centering
\begin{tabular}{|c|cc|}
\hline
& \multicolumn{2}{c|}{Four-fold Knee MRI} \\ \hline
Methods & PSNR & SSIM \\ \hline 
SENSE   & 33.77 ± 1.72 & 0.931 ± 0.019 \\
MoDL    & 37.67 ± 1.61 & 0.969 ± 0.011 \\ 
MOL-SN  & 34.54 ± 1.70 & 0.942 ± 0.017 \\
MOL-L     & 36.38 ± 1.54 & 0.960 ± 0.012 \\
\textbf{MnM-MOL} & \textbf{37.65 ± 1.59} & \textbf{0.968 ± 0.011} \\ 
\hline
\end{tabular}
\vspace{1em}
\caption{Quantitative comparisons on fastMRI knee dataset. Mean PSNR (dB) and mean SSIM are reported.}
\label{tab:fastmri_perf_comp_tab} 
\end{table}

\begin{figure*}[!t]
	\centering
	\includegraphics[width=1.0\textwidth,keepaspectratio=true,trim={0.7cm 3.75cm 0.4cm 4cm},clip]{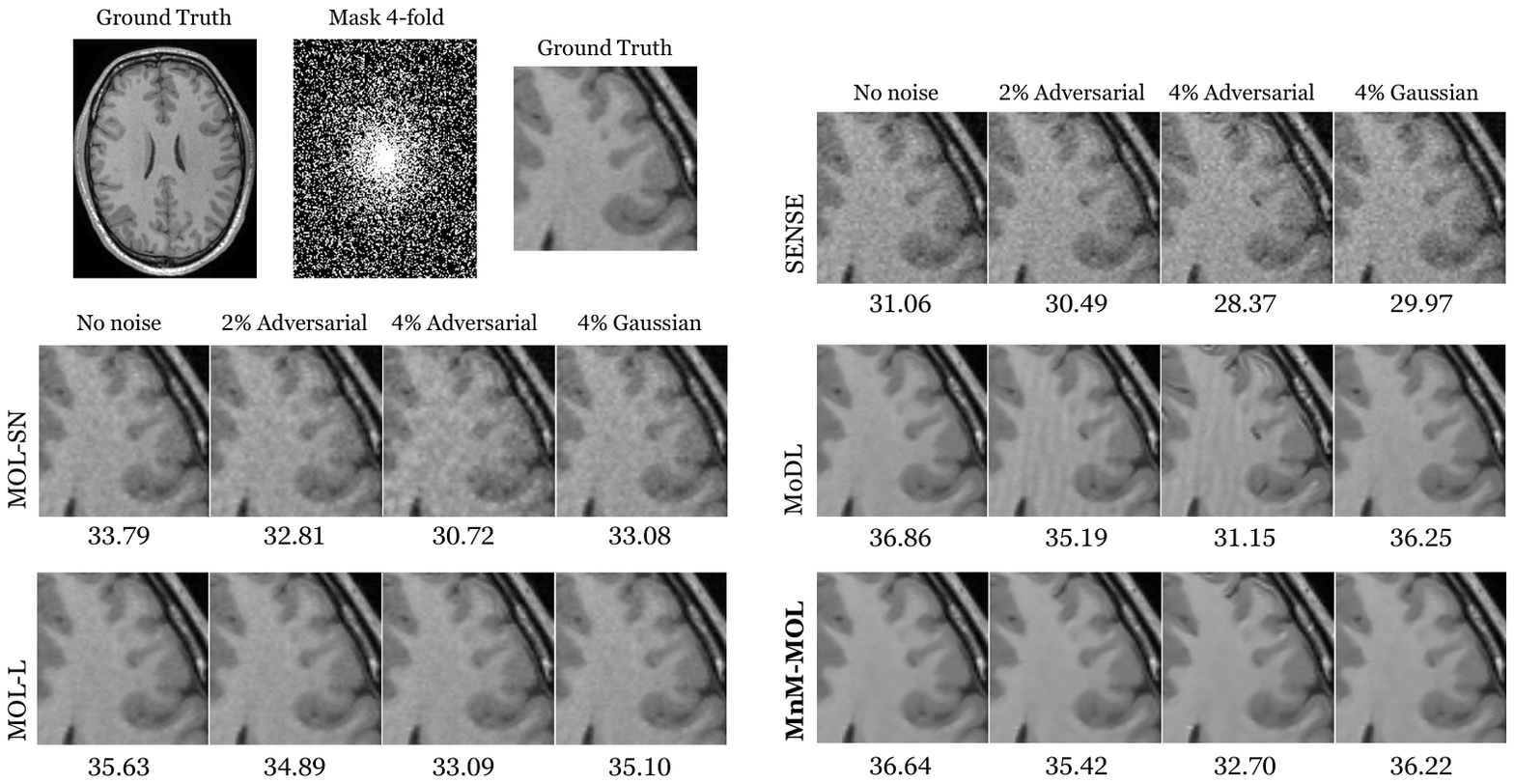}
	\caption{Sensitivity of the algorithms to input perturbations: The rows correspond to reconstructed images of multichannel brain data with 4x acceleration using different methods. The data was undersampled using a Cartesian 2D nonuniform variable-density mask. The columns correspond to recovery without additional noise, with worst-case added adversarial noise  whose norm is 5\% and 10\% of the measured data, and Gaussian noise whose norm is also 10\% of the measured data, respectively. The PSNR (dB) values of the reconstructed images are reported for each case. The improved robustness of the MOL schemes can be attributed to the monotone constraint on the perturbations. We note that the improved performance of MnM-MOL over other MOL schemes is not associated with a loss in robustness. }
	\label{fig:calgary-robustness}
\end{figure*}

The results for four-fold and six-fold accelerated Calgary brain data are shown in Table \ref{tab:calgary_perf_comp_tab} and Fig. \ref{fig:calgary-performance}. Table \ref{tab:calgary_perf_comp_tab} reports the quantitative performance in terms of mean PSNR and SSIM on 20 subjects. The MoDL scheme offers the best PSNR and SSIM measures, which are marginally higher than the proposed MnM-MOL scheme. We also note that the MnM-MOL results appear less noisy in Fig. \ref{fig:calgary-performance}. We observe that the performance of the MOL scheme with spectral normalization (MOL-SN) is significantly worse than MoDL, while switching to a local monotone constraint (MOL-L) improves the performance. We attribute the improved performance of MOL-L over MOL-SN to the relaxation of the global monotone constraint. Likewise, we observe that the use of the MnM scheme further improved the performance over MOL-L. Despite this improvement, the performance of MnM-MOL is marginally lower than that of MoDL. However, we note that all of the MOL approaches are around 10 times more memory-efficient than the unrolled MoDL approach. This improved memory efficiency would enable their application in larger-scale applications (e.g. 3D/4D).  

The results for four-fold accelerated fastMRI knee data are shown in Fig. \ref{fig:fastmri-performance} and Table \ref{tab:fastmri_perf_comp_tab}, respectively. Table \ref{tab:fastmri_perf_comp_tab} reports the quantitative performance in terms of mean PSNR and SSIM on 10 subjects. The results for the fastMRI knee dataset are essentially in line with the results for the brain dataset.

\subsection{Robustness to input perturbations}

We evaluate the robustness of the methods to Gaussian and worst-case input perturbations. We determine the worst-case perturbation $\mathbf n^*$ by solving the optimization problem:
\begin{equation}
	\mathbf n^*= \max_{\mathbf n;~ \|\mathbf n\|_2<\epsilon\cdot\|\mathbf b\|_2} \underbrace{\|\mathbf x^*(\mathbf b + \mathbf n) - \mathbf x^*(\mathbf b)\|_2^2}_{U(\mathbf n)}
\end{equation} 
We solve for $\mathbf n^*$ using a projected gradient ascent algorithm, which alternates between gradient ascent steps and renormalization of $\mathbf n$ to satisfy the constraint $\|\mathbf n\|_2<\epsilon\cdot\|\mathbf b\|_2$.  For MOL-SN, MOL-L, and MnM-MOL, we used fixed-point iterations to compute the gradient. We note that the fixed-point iterations for back-propagation are accurate as long as the forward and backward iterations converge.

\begin{figure}[!h]
	\centering
	\includegraphics[width=0.5\textwidth,keepaspectratio=true,trim={3cm 5.75cm 3cm 6cm},clip]{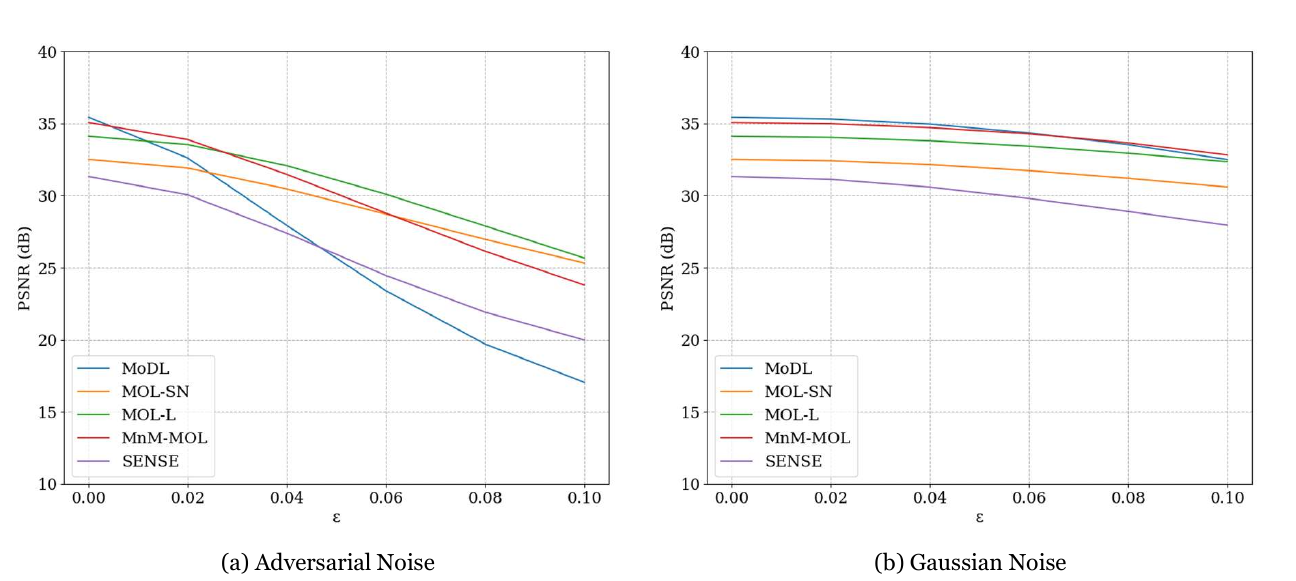}
	\caption{Quantitative comparison of the robustness of different algorithms to worst-case and Gaussian input perturbations for reconstruction of 4x accelerated multi-channel brain data. (a) shows the plot between PSNR and $\delta$, which is the ratio of the norm of perturbations and the norm of the measurements, and (b) shows a similar plot for Gaussian noise.}
	\label{fig:calgary-plots}
\end{figure}

The results for four-fold accelerated brain data are shown in Fig. \ref{fig:calgary-robustness} and Fig. \ref{fig:calgary-plots}. We note that both SENSE and MoDL are more sensitive to adversarial noise. We also note that MoDL is significantly more robust to Gaussian noise than adversarial perturbation. We note that a higher regularization parameter in SENSE would have resulted in a more robust approach, at the expense of performance. Both MOL and MnM-MOL are robust to both adversarial and Gaussian noise. We note that the MnM-MOL scheme can maintain this robustness while offering improved performance over MOL.

\section{Discussion \& Conclusion}

In this paper, we proposed two extensions to the MOL deep equilibrium framework, which constrains the CNN module as a monotone operator. We drew inspiration from CNC methods and constrained the sum of the gradient of the data term and the CNN module to be monotone; this is a relaxation since the CNN module may learn a non-monotone score function, while the sum of the two gradient terms will still be monotone. We replaced the global monotone conditions used in MOL with a more relaxed local monotone constraint to further enhance performance. We also provided novel theoretical guarantees on the uniqueness of the fixed point, convergence, and robustness to input perturbations, when the combined operator is locally monotone and the algorithm is properly initialized.

Our empirical findings demonstrate that when the constraints are relaxed, performance is improved and is comparable to that of unrolled optimization. Because the MOL scheme relies on a DEQ framework, it requires significantly less memory during training compared to unrolled algorithms. Additionally, our experiments on the robustness of the algorithms indicate that the monotone condition leads to better robustness to both Gaussian and adversarial perturbations than unrolled methods.

The radius $\delta$ of the ball $\mathcal B_{\delta}(x); \forall x\in \mathcal M$ is a key factor in the algorithms. To ensure convergence, the iterative algorithm must be initialized with a point inside the ball of radius $\delta$. A bigger $\delta$ will result in a wider basin of convergence, but could translate to more constrained CNN blocks, and hence lower performance. We use fast SENSE reconstructions to initialize the algorithm within $\mathcal B_{\delta}(\mathbf x)$. We note that the performance of SENSE depends on the acceleration factor, and hence a larger basin of attraction (larger $\delta$) is theoretically desired for larger acceleration factors. We also note from Lemma \ref{lem4} that the norm of the perturbations should be less than $m\delta$ for the robustness guarantee to be valid. Therefore, a larger $\delta$ can allow the algorithm to be robust to larger perturbations. Although a larger $\delta$ is desirable for improved convergence and robustness, it may be associated with decreased performance.

\section{Appendix}

\subsection{Proof of Lemma \ref{lem0}}

\begin{proof}
Setting $\mathbf x_{k+1}=\mathbf x_k = \mathbf x^*$ in \eqref{gd_simple}, we obtain $
\mathcal Q_{\theta}(\mathbf x^*) = \mathbf A^H \mathbf b$. Assume that there exist two fixed points $\mathbf x_1^*$ and $\mathbf x_2^*$, then we have $\mathcal Q_{\theta}(\mathbf x_1^*)=\mathbf A^H \mathbf b$ and $\mathcal Q_{\theta}(\mathbf x_2^*) =\mathbf A^H \mathbf b$. Subtracting them, we obtain $\mathcal Q_{\theta}(\mathbf x_1^*)-\mathcal Q_{\theta}(\mathbf x_2^*)= \mathbf 0$. Because $\mathbf x_1^*\neq \mathbf x_2^*$, this is true iff
\begin{eqnarray}\label{zerocondition}
	\Re\Big\langle\mathbf x_1^*-\mathbf x_2^*, \mathcal Q_{\theta}(\mathbf x_1^*)-\mathcal Q_{\theta}(\mathbf x_2^*) \Big\rangle= \mathbf 0
\end{eqnarray}
However, this is not possible when $\mathcal Q_{\theta}$ is an  $m$-monotone operator (see \eqref{eqn:monotonicity_definition}), disproving the assumption.
\end{proof}

\subsection{Proof of Lemma \ref{lem1}}
\begin{proof}
Substituting $\mathcal Q_{\theta} = \mathcal I - \mathcal H_{\theta}$ in the right hand side ($r.h.s = \Re\Big\langle \mathbf z_2-\mathbf z_1, \mathcal \mathcal Q_{\theta}(\mathbf z_2)-\mathcal Q_{\theta}(\mathbf z_1) \Big\rangle$) of \eqref{weaker}, we obtain 
\begin{eqnarray*}\nonumber
 r.h.s &=& 	\|\mathbf z_2-\mathbf z_1\|^2- \\&& \Re\Big\langle \mathbf z_2-\mathbf z_1,\mathcal H_{\theta}(\mathbf z_2)-\mathcal H_{\theta}(\mathbf z_1) \Big\rangle
\end{eqnarray*}
Using Cauchy-Schwarz inequality, we have 
\begin{eqnarray*}
\Re\Big\langle \mathbf z_2-\mathbf z_1,\mathcal H_{\theta}(\mathbf z_2)-\mathcal H_{\theta}(\mathbf z_1) \Big\rangle &\leq& \|\mathbf z_2-\mathbf z_1\|~\|\mathcal H_{\theta}(\mathbf z_2)-\mathcal H_{\theta}(\mathbf z_1)\|\\
&\leq& (1-m) ~\|\mathbf z_2-\mathbf z_1\|^2
\end{eqnarray*}
Here we used the local Lipschitz bound on $\mathcal H_{\theta}$ defined in (\ref{localmonotone}). We thus have 
\begin{eqnarray}
	\Re\Big\langle \mathbf z_2-\mathbf z_1, \mathcal \mathcal Q_{\theta}(\mathbf z_2)-\mathcal Q_{\theta}(\mathbf z_1) \Big\rangle \Big\rangle\geq ~ m~\|\mathbf z_2-\mathbf z_1\|^2.
\end{eqnarray}
which implies that $\mathcal Q_{\theta}$ is $m>0$ locally monotone.
\end{proof}

\subsection{Proof of Lemma \ref{lem3}}
\begin{proof}
 Assume that there exists another fixed point $\mathbf x\neq \mathbf x^*$ within the $\mathcal B_{\delta}(\mathbf x^*)$. Following \eqref{zerocondition}, this is possible iff:
\begin{eqnarray}\label{zerocondition1}
	\left\langle(\mathbf x-\mathbf x^*), \mathcal \mathcal Q_{\theta}(\mathbf x)-\mathcal Q_{\theta}(\mathbf x^*) \right\rangle= \mathbf 0
\end{eqnarray}
However, when $\mathcal Q_{\theta}$ is locally monotone (see \eqref{weaker}), we have:
\begin{eqnarray}
	\left\langle(\mathbf x-\mathbf x^*), \mathcal \mathcal Q_{\theta}(\mathbf x)-\mathcal Q_{\theta}(\mathbf x^*) \right\rangle\geq m\|\mathbf x-\mathbf x^*\|^2,
\end{eqnarray}
which violates \eqref{zerocondition1}. This shows that the fixed point $\mathbf x^*$ is unique within $\mathcal B_{\delta}(\mathbf x^*)$.
\end{proof}

\subsection{Proof of Lemma \ref{lem2}}
\begin{proof}
We first focus on $\gamma = 1/\lambda$, which is easier to analyze and offers the fastest convergence. We note that \eqref{eqn:grad} can be rewritten as
$\lambda\mathcal Q_{\theta}(\mathbf x^*)= \lambda \mathbf A^H \mathbf b$.
Subtracting $\mathbf x^*$ from both sides of \eqref{iterative} and using the above relation, we obtain
\begin{eqnarray}
     \Big(\mathbf x_{k+1}-\mathbf x^*\Big) &=& \mathbf x_k -\mathbf x^* -  ~\mathcal Q_{\theta}(\mathbf x_k)+\mathcal Q_{\theta}(\mathbf x^*)\\
     &=& \mathcal H_{\theta}\left(\mathbf x_k\right)-\mathcal H_{\theta}(\mathbf x^*)
\end{eqnarray}
Note that $\|\mathbf x_0-\mathbf x^*\|<\delta$. Using the Lipschitz  bound of $\mathcal H_{\theta}$, we thus have 
\begin{eqnarray}
     \|\mathbf x_{1}-\mathbf x^*\|
     &<& (1-m)~\delta
\end{eqnarray}
When $(1-m)<1$, we have $\mathbf x_1 \in \mathcal B_{\delta}(\mathbf x^*)$. We thus see that 
\begin{eqnarray}
     \|\mathbf x_{k+1}-\mathbf x^*\|
     &<& (1-m)\|\mathbf x_k-\mathbf x^*\|\\
     &<& (1-m)^k \|\mathbf x_0-\mathbf x^*\|\\
     &<& (1-m)^k \delta, 
\end{eqnarray}
which implies that $\|\mathbf x_{k+1}-\mathbf x^*\| \rightarrow 0$ as $k\rightarrow \infty$.

When $\gamma \lambda < 1$, we have
\begin{eqnarray*}
     \Big(\mathbf x_{k+1}-\mathbf x^*\Big) &=& \Big(\mathbf x_k -\mathbf x^*\Big) -  \gamma\lambda ~\left(\mathcal Q_{\theta}(\mathbf x_k)-\mathcal Q_{\theta}(\mathbf x^*)\right)\\
     &=& \Big(\mathbf x_k -\mathbf x^*\Big)(1-\gamma\lambda)\\
     &&\qquad+\gamma\lambda ~\left(\mathcal H_{\theta}(\mathbf x_k)-\mathcal H_{\theta}(\mathbf x^*)\right)
\end{eqnarray*}
Using triangle inequality and \eqref{localmonotone}, we have 
\begin{eqnarray*}
     \|\mathbf x_{k+1}-\mathbf x^*\| &\leq& 
\|\mathbf x_k -\mathbf x^*\|\Big(1-\gamma \lambda m\Big)
\end{eqnarray*}
Because $(1-\gamma \lambda m)<1$, and using similar arguments as before, we have $\|\mathbf x_k -\mathbf x^*\|\rightarrow 0$ as $k\rightarrow \infty$.
\end{proof}

\subsection{Proof of Lemma \ref{lem4}}
\begin{proof}
We rewrite the iterative rule in \eqref{gd_simple} as an expanded fashion as:
\begin{eqnarray*}
\mathbf x_{k} &=& \mathcal H_{\theta}^k(\mathbf x_{k-1})+ \mathbf A^H \mathbf b\\ 
&=&\mathcal H_{\theta}^k(\mathbf x_0) + \Big(\mathcal H_{\theta}^{k-1}+\mathcal H_{\theta}^{k-2}+\ldots \mathcal H_{\theta} +\mathcal I\Big)\left( \mathbf A^H \mathbf b\right)
\end{eqnarray*}
When the measurements $\mathbf b$ are perturbed, the iterations are specified by
\begin{eqnarray*}
\mathbf y_{k} &=& \mathcal H_{\theta}^k(\mathbf y_{k-1})+ \mathbf A^H \mathbf b+ \mathbf A^H \mathbf n\\ 
&=&\mathcal H_{\theta}^k(\mathbf x_0) + \Big(\mathcal H_{\theta}^{k-1}+\mathcal H_{\theta}^{k-2}+\ldots \mathcal H_{\theta} +\mathcal I\Big)\left( \mathbf A^T \mathbf b\right)\\&&+\Big(\mathcal H_{\theta}^{k-2}+\ldots \mathcal H_{\theta} +\mathcal I\Big)\left( \mathbf A^H \mathbf n\right)
\end{eqnarray*}

The difference between the $k^{\rm th}$ iterates is given by 
\begin{eqnarray*}
\mathbf y_{k}-\mathbf x_{k} &=& \mathcal H_{\theta}^k(\mathbf y_{0})-\mathcal H_{\theta}^k(\mathbf x_{0}) +\\&& \Big(\mathcal H_{\theta}^{k-1}+\mathcal H_{\theta}^{k-2}+\ldots \mathcal H_{\theta} +\mathcal I\Big) \mathbf A^H\mathbf n
\end{eqnarray*}
Using triangle inequality and denoting the local Lipschitz constant of $\mathcal H_{\theta}$ by $L=1-m$, we obtain
\begin{eqnarray*}
\|\mathbf y_{k}-\mathbf x_{k}\| &=& L^k\|\mathbf y_{0}-\mathbf x_{0}\| +\\&& \Big(L^{k-1}+L^{k-2}+\ldots L +1\Big)\|\mathbf A^H\mathbf n\|
\end{eqnarray*}
As $k\rightarrow \infty$, the first term vanishes because $L<1$. Therefore, we have 
\begin{eqnarray*}
\|\Delta\|=\|\mathbf y^*-\mathbf x^*\| &=& \frac{\|\mathbf A^H\mathbf n\|}{\Big(1-L\Big)} = \frac{\|\mathbf A^H\mathbf n\|}{m}
\end{eqnarray*}
All of the above arguments are valid only if $\mathbf y_{\infty}$ as well as $\mathbf y_k$ are still within the $\mathcal B_{\delta}(\mathbf x)$. Setting $\|\Delta\| \leq \delta$, we obtain the condition $\|\mathbf A^H\mathbf n\|\leq m\delta$.
\end{proof}

\bibliographystyle{IEEEtran}
\bibliography{ref}
\end{document}